\documentstyle[psfig]{article} 
\begin{document}

\centerline{\Large \bf
NASA's Future Missions in X-ray Astronomy
}  

\bigskip
\centerline{\bf Nicholas E. White}
\centerline {NASA's Goddard Space Flight Center}
\centerline{Laboratory for High Energy Astrophysics}
\centerline {Greenbelt, MD 20771 USA}
\centerline {nwhite@lheapop.gsfc.nasa.gov}
 
\bigskip
\section*{Abstract}

The NASA program in X-ray astronomy has two long term goals: 1) to
achieve sufficient angular resolution to image the event horizon of a
black hole (0.1 micro arc sec) and 2) to achieve sufficient
collecting area (50-150 sq m) and angular resolution (0.1--1.0 arc sec) to
observe in detail the first black holes and galaxies at high
redshift. These ambitous goals can be used to map out a series of
missions and a technology program. The next major mission will be 
Constellation-X which will be dedicated to high resolution X-ray spectroscopy for launch in $\sim$ 2010. This mission is a critical step in the roadmap to achieve these goals. Following Constellation-X NASA is considering two very ambitious {\it vision} missions: MAXIM and Generation-X that will achieve the ultimate capabilities. The modest missions Astro-E2 and Swift address more focussed science goals on a rapid development cycle and provide important pathfinders to the larger missions.

\medskip 
{\it Key-words}{ X-ray Astronomy, Missions}

\section{Introduction}

The current generation of large X-ray observatories: the Chandra X-ray Observatory from NASA and the XMM-Newton Observatory from ESA are delivering an impressive array of results. As we look to the future the requirements for future X-ray observatories are driven by some basic questions: What is the nature of space and time? When did the first black holes appear? What is the Universe made of? X-ray observations are central to addressing these questions. Using these questions we can plan a roadmap for the future of X-ray astronomy.

\subsection{What is the nature of Space and Time?}

Black holes represent the most extreme environments known. They provide the only known regime where the effects of General Relativity (GR) can be tested in the strong gravity regime. Material falling into the black hole is heated to high temperatures and the bulk of the energy is emitted in X-rays. Results from the Japanese-US ASCA mission have demonstrated that X-ray spectral features are coming from close to the event horizon of the black hole. These results suggest that X-ray observations hold the promise of using black holes to test GR. The most direct way to do this is to resolve  the event horizon of a black hole and directly observe the predicted distortions of space time. The best candidates for this are the super massive black holes in nearby active galactic nuclei, such as M87. To resolve the event horizon will require a resolution of 0.1 to 1.0 micro arc sec. This can only be achieved with X-ray interferometry and is a similar challenge in scope to the plans for the Terrestrial Planet Finder (TPF) and Darwin missions. The MAXIM mission will address this science goal. In the nearer term, the study of spectroscopic line signatures from close to the event horizon requires an 100 times increased sensitivity compared to Chandra and XMM-Newton for high resolution spectroscopy with a broad 0.25 to 40 keV energy range.  Such a capability is more accessible with technology that is an extension of the current generation of missions and will be addressed by the Constellation-X mission. 

\subsection{When did the first black holes appear?}

Current theories predict that the first black holes formed at redshifts of 10-20 with masses of 10$^3$ to 10$^6$ solar masses and that they play a crucial role in galaxy formation. X-ray observations are essential to observe these black holes turn on as proto-quasars, and evolve with cosmic time. The Chandra and XMM-Newton deep surveys are revealing that active super-massive ($>$ 1 million solar mass) black holes are present in up to 10\% of galaxies at redshifts of 1-3 (compared to 1\% in the local universe). These results are demonstrating the power of X-ray surveys to find accreting black holes in otherwise normal galaxies. X-ray spectroscopic observations can be used to determine the evolution of the black hole spin with cosmic time, and infer black hole mass from reverberation analysis and X-ray luminosity scaling arguments. To observe the first black holes requires a collecting area 100 to 1,000 times Chandra and XMM-Newton, with $\sim$  0.1--1 arc sec angular resolution. This capability would also allow the X-ray detection of normal galaxies essentially throughout the Universe, which will be important to determine whether black holes formed, before, during, or after, the era of galaxy formation. The Generation-X mission is tuned to directly achieve this goal. The nearer term Constellation-X mission will provide critical measurements of the evolution of active galactic nuclei with redshift.

\subsection{What is the Universe made of?}

A pressing challenge in astronomy today is determining the composition of the Universe. Using clusters of galaxies as cosmological probes, X-ray observations have constrained the matter content (dark and baryonic) to be about 35\% of the closure density. This result was a precursor to the recent revelation from observations of the cosmic microwave background and type 1a supernovae that the remaining 65\% of the universe is made up of a mysterious dark energy. X-ray observations are central to this problem because the soft and medium X-ray band contains spectral diagnostics from all the abundant elements formed after the big bang. X-ray observations of supernovae, the hot interstellar and intergalactic medium, galaxies and clusters of galaxies provide a direct probe of when the elements from which we are made were created and the formation of large scale structure. Current cosmological models predict that most of the baryons in the Universe are in a hot intergalactic medium, and that sufficiently sensitive X-ray observations will detect this gas in the absorption spectra of high redshift quasars, or other bright distant objects. This requires an increase in sensitivity by a factor of 100 relative to Chandra and XMM-Newton for high resolution (R$ > $300) spectroscopy. The same increase in capability will also allow precise measurements of the first clusters of galaxies at high redshift which, in concert with future observations of the microwave background and of type Ia supernovae, will tightly constrain the overall energy and mass content of the universe. In addition, an extension of imaging observations to the hard X-ray band will be very important to investigate the contribution of non-thermal processes. Constellation-X will directly address this question via high resolution spectroscopy.

\section{Constellation-X}

The Constellation-X mission is a large collecting area X-ray observatory emphasizing high spectral resolution (R $\sim$ 300 to 3000) and a broad energy band (0.25 to 60 keV). It is the next major mission being planned by NASA for X-ray astronomy. Constellation-X is a high priority in the US National Academy of Sciences McKee-Taylor 2000-2010 Survey for a new start and flight in the first decade of the 21st century. 

The 0.25--10 keV X-ray band contains the K-shell lines for all of the abundant metals (carbon through zinc), as well as many of the L-shell lines.  The detailed X-ray line spectra are rich in plasma diagnostics which also provide unambiguous constraints on physical conditions in the sources. A spectral resolving power of at least 300 is required to separate the He-like, density sensitive, triplet from ions such as O, Si, and S. In the region near the iron K complex a resolving power exceeding 2000 is necessary to distinguish the
lithium-like satellite lines from the overlapping helium-like transitions. A resolution of 300--3000 provides absolute velocity measurements of 10--100 km/s, commonly found in many astrophysical settings. 

By increasing the telescope aperture to 3 m$^2$ with 5--15 arc sec resolution and utilizing efficient spectrometers the mission will achieve the required 25 to 100 increased sensitivity over current high resolution X-ray spectroscopy missions. The use of focusing optics across the 10 to 60 keV band will provide a similar factor of 100 increased sensitivity in this band, essential for constraining the nature of the underlying continuum. The effective
area curves of Constellation-X are shown in Figure 1, and compared with
the equivalent curves from the high resolution spectrometers on {\it Chandra}
and XMM-Newton. The peak telescope area of 
30,000 cm$^2$ when combined with
the efficiency of the spectrometers and detectors gives the required 
15,000 cm$^2$ at 1 keV and 6,000 cm$^2$ at 6 keV. The effective area of 
1,500 cm$^2$ at 40 keV is set by using a typical AGN spectrum to match the 
sensitivity in the 0.25--10 keV band. A $10^5$ s exposure of a typical AGN 
with the Constellation-X will 
obtain $\sim$ 1000 counts in the calorimeter at 
a flux of $\sim$ $2 \times 10^{-15}$ erg cm$^{-2}$ s$^{-1}$ (0.2--2 
keV band). 

\begin{figure}
\centerline{\hbox{
\psfig{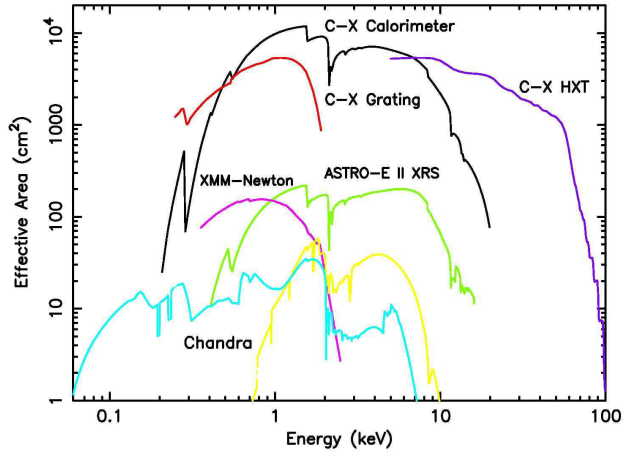}
}}
\caption{The effective area vs. energy for the three Constellation-X
instruments (co-aligned and operating simultaneously) for the
four-satellite system. For comparison the high resolution
spectrometers on {\it Chandra LETG (solid) and HEG (light)}, {\it XMM-Newton} (RGS) and Astro-E2 XRS are shown.}
\end{figure}

A spectroscopy X-ray telescope (SXT) covers the 0.25--10 keV band. 
X-ray spectroscopy requires that a maximum number of photons be collected.
To maximize the collecting area per unit mass an angular
resolution requirement of $15^{''}$ half power diameter (HPD) is set for the SXT as the 
baseline requirement, with a goal of $5^{''}$. The $15^{''}$ HPD requirement 
is consistent with the confusion limit for the limiting flux of the sources 
to be studied. The SXT uses two complementary spectrometer systems to achieve 
the desired energy resolution of 300--3000:  an array of high 
efficiency quantum micro-calorimeters with energy resolution of 2eV,
and a set of reflection gratings with a resolution of 0.05$\AA$ in the first 
order. The gratings deflect part of the telescope beam away from the
calorimeter array to a CCD array in a design similar to {\it XMM-Newton}, except that the
Constellation-X direct beam falls on a high spectral resolution
quantum calorimeter instead of on a CCD. The baseline field of view of the 
calorimeter is $2.5^{'}$ square, with a 30 x 30 array. This gives 
$5^{''}$ pixels, sized to adequately sample the telescope point spread function.

The Constellation-X hard X-ray telescope (HXT) baseline design uses 
multi-layers to provide a focusing optics system that operates in the band above 10 keV band. The 
baseline upper energy is 40 keV, with a goal of reaching 60 keV. The improvement 
in the signal to noise results in a factor of 100 or more increased sensitivity
over non-focussing methods used in this band.  
The requirement for the HXT angular resolution is $ 1^{'}$ HPD (with a goal 
of $30^{''}$) which is quite sufficient to resolve sources in this energy band. The 
FOV is $8^{'}$, or larger. There are no strong atomic lines expected above 10 keV, so there is a 
relatively modest HXT spectral resolution requirement of R $>$10 across the band.

The required large collecting area is achieved 
utilizing several mirror modules, each with its own
spectrometer/detector system. The Constellation-X design  
takes advantage of the fact that several smaller 
spacecraft and more modest launch vehicles each carrying
one ``science unit'' can cost less than one very large
spacecraft and launcher. The program is then very robust in 
that risks are distributed over several launches and 
spacecraft with no single failure leading to loss of mission. 
The current baseline mission is four satellites that are carried in pairs on 
either two Atlas V or Delta IV launchers. 
The interval between the two launches will be 
$\sim$ 1 year, with the first launch in the 2010 timeframe. 
A L2 orbit will facilitate high observing efficiency, provide an 
environment optimal for cyrogenic cooling, and simplify the spacecraft design. 
The telescope collecting area is sized assuming $\sim$ 95\% observing 
efficiency, which maximizes the observing 
program for a minimum 5 year mission lifetime (with all satellites operating).







\section{Vision Missions}

To set the long term goals, NASA has put on the road map two "vision missions" for the $>$2015 timeframe

\subsection{Generation-X}

Generation-X is designed to observe the high redshift (5-20) Universe to observe X-ray emission from the first black holes and starburst galaxies at redshift 5-20, and to observe the X-ray evolution of galaxies over all redshifts. These science objectives require that the mission has 50-150 m$^2$ collecting area at 1 keV combined with Chandra-like angular resolution of 0.1 to 1.0 arc sec resolution. This resolution is necessary to provide the necessary detection sensitivity and avoid source confusion. The large collecting area would be achieved by developing precision light weight X-ray optics, with a mass per unit collecting area similar to the Astro-E optics, but with a factor of 100 improved angular resolution. As currently envisioned the mission would consist of 3-6 identical satellites, each carrying a fraction of the collecting area. Deployable optics would be used to maximize the collecting area per launch vehicle. The focal length of 25 to 50 m would be achieved by using an extendible mast. The mission would be placed at L2 to provide a stable thermal environment and to maximize the viewing efficiency. The enabling technology for this mission is lightweight, precision, grazing incidence optics and the mission requirements feed into NASA's long term technology program.

\subsection{MAXIM}

The Micro-Arcsecond X-ray Imaging Mission (MAXIM) will have the angular resolution of  0.1 to 1 micro arc second required to resolve the event horizon of accreting black holes at the center of nearby galaxies (e.g. M87). MAXIM will employ for the first time X-ray interferometry and in doing so will achieve a spectacular 10 million times increase in angular resolution compared to Chandra. The basic technique has recently been demonstrated in the laboratory. To obtain 0.1 micro arcsec requires a 100 m to 1000 m baseline. The mission is being studied by NASA's Institute for Advanced Concepts (NIAC). It involves up to 34 spacecraft flying in formation, with precision metrology and tolerances at the nano-meter level. On orbit metrology and stability requirements are similar to TPF/Darwin.  A two spacecraft Pathfinder mission is being studied that will have a 1 m baseline giving 0.1 milli-arc second imaging. This will provide the required demonstration of X-ray interferometer in space and at the same time give a 1,000 times improvement compared to Chandra, which in itself will allow exciting new science e.g. imaging coronal structures on nearby stars and the base of jets in active galactic nuclei. The full MAXIM is still very early in a concept study phase and the technology requirements are very challenging.

\section{Modest Missions}

It is also important that there is an adequate number of more modest missions that focus on more specialized scientific goals. NASA pursues these small to medium sized opportunities via its Explorer program, which allows rapid response to topical science challenges.

\subsection{Swift}

Swift is a multiwavelength observatory for gamma-ray burst (GRB) astronomy, that is in development for launch in 2003.  It is the first-of-its-kind multi-wavelength, autonomous rapid slewing satellite for studying explosive astronomical events.  It will be far more powerful than any previous GRB mission, observing 150-300 bursts per year and performing detailed X-ray and UV/optical afterglow observations spanning timescales from 1 minute to several days after the burst. The mission is being developed by an international collaboration led by NASA, with Italy and the UK providing significant parts of the science payload and operations. The mission will carry three instruments: a new-generation wide-field gamma-ray detector and narrow-field X-ray and optical telescopes. Innovations from the Swift program include: 1) a large-area gamma-ray detector using CdZnTe detectors; 2) an autonomous rapid slewing spacecraft; 3) a multi-wavelength payload combining optical, X-ray, and gamma-ray instruments; 4) immediate data flow to the community.

\subsection{Astro-E2}

The Astro-E2 mission is a Japanese mission, with substantial US participation. It is a re-flight of Astro-E, which suffered a launcher failure on February 10, 2000.  Astro-E2 is approved to be launched with an ISAS M-V rocket in early 2005. Astro-E2 will basically be identical to Astro-E.  The spacecraft will carry five X-Ray Telescopes (XRTs). One telescope module will feed the X-ray Spectrometer (XRS) and the other four telescopes will have X-ray Imaging Spectrometer (XIS) units at their focal planes.  At the same time, the Hard X-ray Detector (HXD) will be able to measure high-energy X-rays.  The development of the XRT, XRS and XIS are being performed in collaboration with NASA. Even though Astro-E2 will be realized about 5 years after Astro-E was expected these capabilities remain unsurpassed. The lifetime of the XRS will be limited to 2--3 yr by expendable cryogenics. The broad band-pass and  spectroscopic response will provide a unique capability for addressing a wide variety of fundamental problems in astrophysics including the origin of the elements and structures in the Universe as well as the evolution and dynamics of these structures.

\end{document}